# Resonant THz sensor for paper quality monitoring using THz fiber Bragg gratings

Guofeng Yan,[1,3] Andrey Markov,[1] Yasser Chinifooroshan,[2] Saurabh M. Tripathi,[2] Wojtek J. Bock,[2] and Maksim Skorobogatiy[1,*]

[1] *Department of Engineering Physics, Ecole Polytechnique de Montréal, Québec, Canada*
[2] *Département d'informatique et d'ingénierie, Université du Québec en Outaouais, Québec, Canada*
[3] *Centre for Optical and Electromagnetic Research, Zhejiang University, P. R. China*
*Corresponding author: maksim.skorobogatiy@polymtl.ca*

*We report fabrication of THz fiber Bragg gratings (TFBG) using $CO_2$ laser inscription on subwavelength step-index polymer fibers. A fiber Bragg grating with 48 periods features a ~4 GHz-wide stop band and ~15 dB transmission loss in the middle of a stop band. The potential of such gratings in design of resonant sensor for monitoring of paper quality is demonstrated. Experimental spectral sensitivity of the TFBG-based paper thickness sensor was found to be ~ -0.67 GHz / 10 μm. A 3D electromagnetic model of a Bragg grating was used to explain experimental findings.*

Terahertz (THz) waves offer unique opportunities that are not available at other wavelengths. Firstly, THz radiation is non-ionizing, which can be employed for safe imaging. Secondly, many materials, such as ceramic [1], plastic [2], and paper-based materials [3] are relatively transparent for THz radiation. Hence, innovative security and quality-control applications are envisioned by utilizing THz waves. Most of the current THz sensors are realized in the non-resonant configurations where sample is interrogated directly by the THz light. In resonant sensors, changes in the sample properties are measured indirectly by studying variations in the optical properties of a resonant structure coupled to a sample. In this letter, we study the use of THz fiber Bragg grating for resonant THz sensing.

Fabrication of the THz fiber Bragg gratings (TFBGs) requires availability of the low-loss waveguides. One of the simplest examples of a low-loss THz waveguide is a subwavelength step-index plastic fiber [4]. In such fibers, a large fraction of the modal power is found outside of the fiber core, which not only significantly reduces the transmission losses, but also makes such fibers a promising platform for sensing applications. Recently, fabrication of several fiber-based components such as TFBGs [5] and THz notch filters [6] was reported using laser inscription on THz fibers. Very precise, however, expensive excimer laser system was used for inscription.

In this work, we demonstrate fabrication of TFBGs using a cost effective $CO_2$ laser inscription system, which is widely used for long period grating fabrication on silica fibers in the near-IR [7, 8]. Optical response of such gratings and their application in sensing were studied theoretically using GratingMOD module from RSoft Design Group. Fabricated TFBGs were then used to build prototypes of resonant multi-measurand sensors for monitoring paper quality, which is an important manufacturing problem [9]. For the lack of space, in this letter we only present the use of such sensors for paper thickness monitoring, while only indicating the other sensing modalities.

The subwavelength step-index fibers with a diameter of 400 μm were drawn in-house from the low density polystyrene (LDPE) rods. LDPE has a relatively low material loss (~0.2 $cm^{-1}$) and almost constant refractive index (RI) of 1.54 in the 0.2-0.5 THz region studied in this work [4]. TFBGs were then inscribed point-by-point on the 5 cm-long pieces of step-index LDPE fibers using a class IV Synrad $CO_2$ laser operating at 10.6 μm with average output power of 1.5 W and repetition rate of 20 kHz. The TFBG presented in this report consists of 48 notches that are ~110 μm-deep and ~170 μm-wide (see Fig. 1 (a)). Grating period is designed to be 340 μm, in order to place the TFBG stop band in the low-loss transmission window of the subwavelength fiber which is centered at ~0.3 THz. TFBG total length is 17 mm.

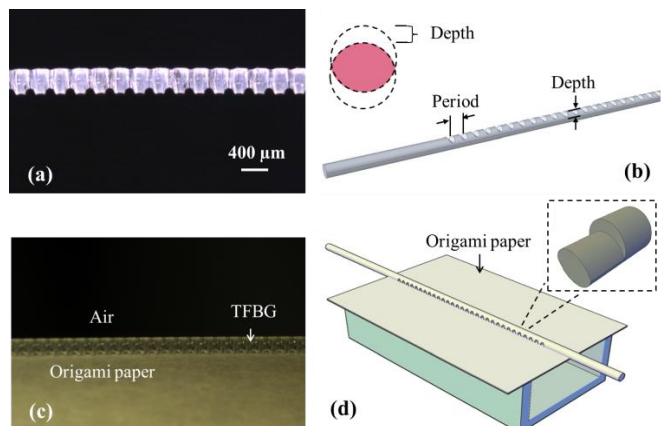

Fig. 1. Microscope images of the fabricated TFBG (a), and of the paper thickness monitoring setup (c). Schematics of the TFBG 3D model (b), and of the paper thickness monitoring.

The TFBG transmission spectra were recorded using a THz-time domain spectroscopy (TDS) setup modified for fiber measurements [10]. In our studies we used ~600 ps-long scans that resulted in spectral resolution of ~1.5 GHz. A ~4 GHz-wide stop band (Full Width at Half Maximum,

FWHM) at ~375 GHz and ~15 dB transmission loss in the middle of a stop band were observed (see Fig. 2). Similar to the fiber Bragg gratings in the near-IR, spectral position and shape of the TFBG transmission peak are sensitive to changes in the environment, which opens an opportunity for their use in sensing and monitoring.

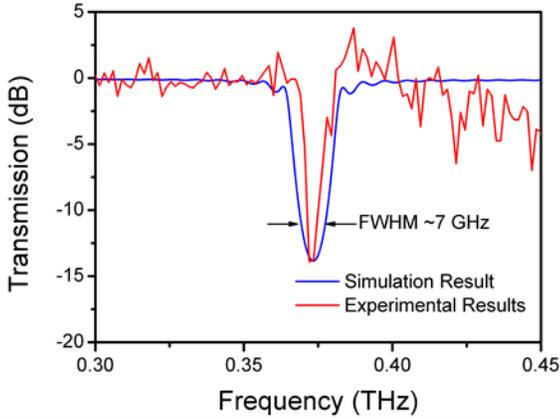

Fig. 2. Comparison between simulated and experimental transmission spectra of TFBG.

In our simulations, TFBG transmission spectra were modeled using the GratingMOD (RSoft Design Group) software module, which is based on the Coupled-Mode Theory and the transfer matrix method [11]. The geometrical model of the TFBG is simply a solid cylinder with multiple annular indentations that represent notches (see Fig. 1 (b)). The cross section of the notches is shown in Fig.1 (b), which is an intersection of the two rods (black dash curves) with diameters of 400 µm and center-to-center distance defined as depth of the notch. An enlarged 3D geometrical model of one period is also given in Fig. 1(d). The geometrical parameters such as width and depth of the notches were measured from the Bragg grating micrographs like the one presented in Fig. 1(a). Forty-eight grating periods were simulated with a period of 340 µm, notch depth of 110 µm and notch width of 170 µm were simulated. The transmission spectrum of the TFBG without paper layer is shown in Fig. 2. When compared to the experimental results, the transmission loss and the dip position of the experimental results fit very well with the simulated ones. The width of the TFBG peak is somewhat larger (~7 GHz) than the experimental one, which is largely due to geometrical differences between simulated and experimental notch profiles.

Next, we have measured frequency dependent refractive index and absorption coefficient of the Origami paper by using the same THz TDS setup. Similarly to the cut-back method, transmission measurements through the stacks of 2, 4, 6, and 8 layers of 60 µm-thick Origami paper were carried out and the paper absorption loss and refractive index were extracted from this data. The results are shown in Fig. 3. The refractive index of the Origami paper (blue circles) is relatively constant ~1.43 in the 0.2-0.4 THz frequency range, which is consistent with the measurements in [9]. The paper absorption coefficient (red squares in Fig. 3), however, shows rapid increase at higher frequencies from ~5 cm$^{-1}$ a 0.2 THz to ~20 cm$^{-1}$ at 0.6 THz. In the frequency range of a TFBG stop band 0.3 - 0.4 THz, the absorption coefficient is ~ 4 cm$^{-1}$.

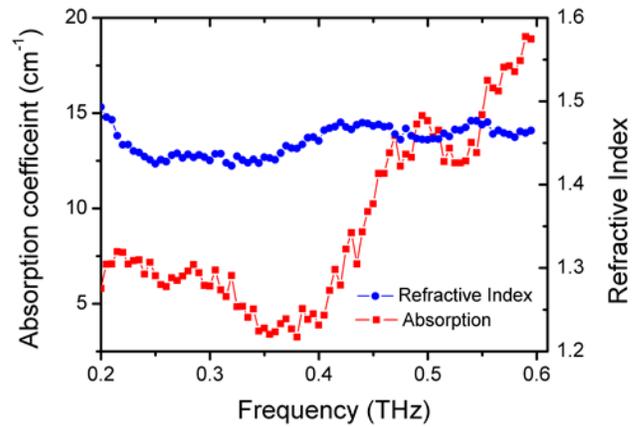

Fig. 3. Frequency dependence of the refractive index (blue circles) and absorption coefficients (red squares) of the Origami paper.

We now study changes in the TFBG transmission spectrum in response to changes in the thickness of a paper stack, which is brought in contact with the grating (see Fig. 1(d)). In our simulations we use experimentally measured values of the paper refractive index, absorption coefficient and paper thickness to obtain the TFBG transmission curves presented in Fig. 4. We note that TFBG transmission dip shifts towards lower frequencies from 373.8 GHz to 356.3 GHz (longer wavelengths) as the paper stack thickness increases from 60 µm to 240 µm. This behavior is a simple manifestation of increase in the TFBG modal effective refractive index $n_{eff}$ due to presence of paper in the grating vicinity. As higher modal refractive index results in longer Bragg wavelength as

$$\lambda_B = 2n_{eff}\Lambda, \quad (1)$$

hence, the grating peak shifts to lower frequencies for thicker paper stacks.

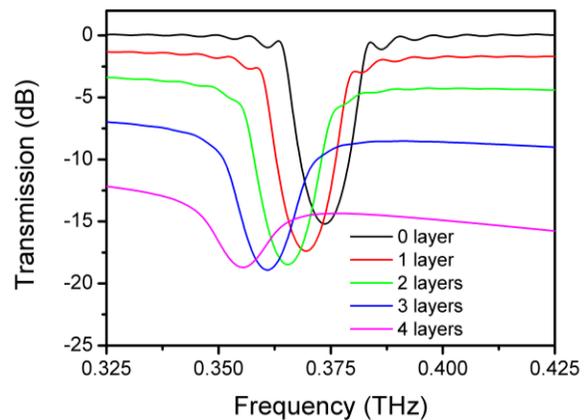

Fig. 4 Simulated TFBG transmission spectra for a different number of paper layers placed in direct contact with TFBG.

From results in Fig. 4 we can estimate the spectral sensitivity of our sensor to changes in the paper thickness as -0.73 GHz/10 µm. Given that spectral resolution of our

THz TDS setup is 1.5 GHz, then, the minimal measurable change in the paper thickens is expected to be ~ 20 μm.

Additionally, it is clear from Fig. 4 that outside of the TFBG stop band, the use of thicker paper stacks results in higher transmission losses. This finding is simple to rationalize, as in the presence of thicker stacks, the guided grating mode will have stronger presence in the lossy paper region. To make this statement more evident, in Fig 5 (a) we present field distributions of the fundamental mode of a subwavelength fiber coupled to paper stacks of increasing thickness. The fiber is uniform along its length and it has the same diameter of 400 μm as the one used in TFBG fabrication. Clearly, when increasing the thickness of a paper stack, the fundamental guided mode features stronger localization in the paper region, and, as a consequence, higher effective refractive index (see Fig.5 (b)) and higher losses (see Fig. 5(c)).

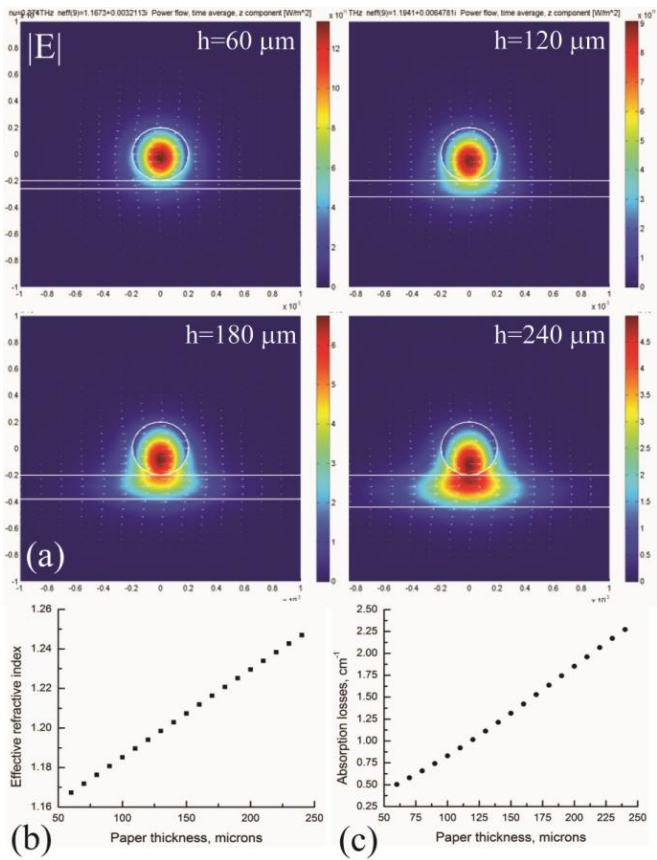

Fig. 5 (a) Field distribution in the fundamental mode of a subwavelength fiber coupled to paper stacks of increasing thickness. (b) Modal effective refractive index and (c) modal loss as a function of the paper stack thickness.

In principle, amplitude changes in the TFBG transmission losses outside of the stop band can be used as an alternative or supplementary method for monitoring paper thickness. Thus, from Fig. 4 we can deduce that at 0.375 THz, amplitude sensitivity to changes in the paper thickness is ~ 0.8 dB / 10 μm, which is quite high. In practice, however, we find that amplitude measurement with our THz setup is considerably more sensitive to noise than a spectral measurement, thus making amplitude detection a considerably less sensitive and less attractive alternative.

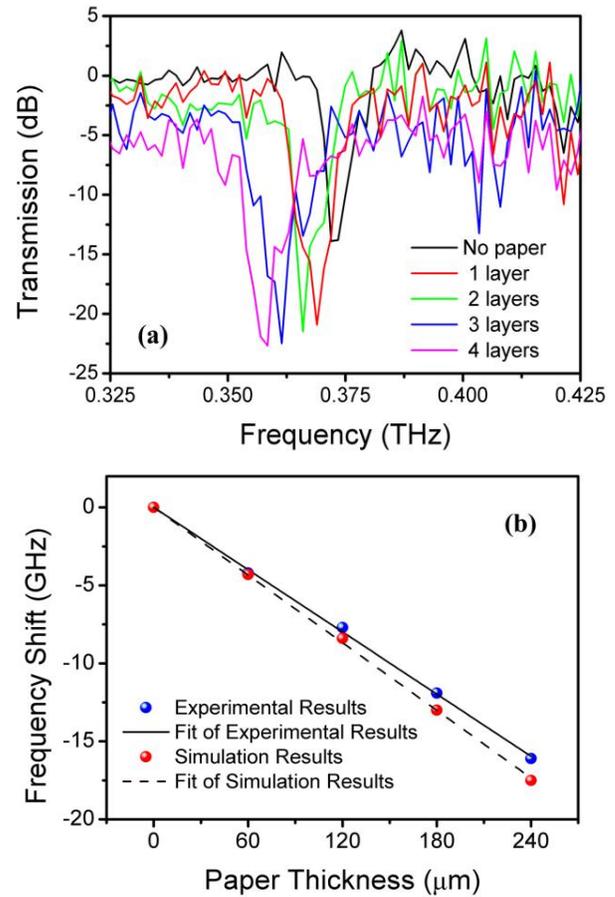

Fig. 6 (a) Experimental TFBG transmission spectra for a different number of paper layers placed in direct contact with TFBG. (b) Comparison of the simulated and experimental frequency shifts of the TFBG peak position as a function of the total paper thickness.

To test our numerical predictions, experimental measurements of the TFBG transmission spectra have been performed for various number of paper layers placed in direct contact with a TFBG (see Fig. 1(b), (c)). The paper sheets were of the same width as the grating region (17 mm). By placing more layers of the Origami paper in contact with a TFBG, position of the transmission dip shifted proportionally to the paper thickness from 373.4 GHz to 357.3 GHz (see Fig. 6(a)). Comparisons between simulation and experimental results show very good correspondence in the center peak position (see Fig. 6(b)). Experimental spectral sensitivity of the peak position to changes in the paper thickness is found to be -0.67 GHz / 10 μm, which agrees very well with the simulated one - 0.73 GHz / 10 μm. By comparing Fig. 6(a) with Fig. 4 we note that measured transmission losses are somewhat smaller than the simulated ones. We attribute the lower measured losses to the fact that paper layers in the stack are not tightly pressed, thus allowing some low-loss air regions between the paper layers. Finally, using the measured sensitivity value we conclude that with a current THz setup that offers 1.5 GHz resolution, we can reliably detect ~20 μm variations in the paper thickness.

We believe that such sensitivities are high enough to be of interest for applications in real-time monitoring of paper manufacturing lines.

At this point, it is timely to summarize and compare various sensing modalities offered by the TFBGs. Firstly, TFBG-based sensors can be operated using spectral interrogation. In this case, changes in the paper thickness are deduced directly from the spectral shifts of the grating stop band. Importantly, paper loss does not contribute to the spectral position of a stop band. Experimentally, we also find that spectral interrogation is much less sensitive to noise compared to the amplitude-based interrogation, with the sensor resolution limited mainly by the spectral resolution of a THz TDS setup.

Secondly, changes in the paper thickness can be detected by monitoring transmission losses outside of the TFBG bandgap (see, for example, transmission amplitudes at 0.325 THz in Fig. 4, and Fig. 6(a)). This is, essentially, an amplitude-based interrogation method, with thicker paper stacks resulting in the higher TFBG transmission loss. This method, however, requires using paper with relatively high absorption loss in order to enhance the effect of increased modal presence in thicker paper stacks on TFBG transmission loss. Moreover, TFBG loss is sensitive to changes both in the paper thickness and in the paper humidity; therefore, an amplitude-based interrogation will probe the net effect of variations in both parameters. In principle, using together the spectral-based and amplitude-based methods allows deconvolution of the two measurands (paper thickness and paper humidity), however, a relatively high noise level in the experimentally measured transmission spectra (see Fig. 6(a)) prevented us from attempting such a deconvolution.

Thirdly, analysis of changes in the phase of transmitted light can give yet another modality for the multi-measurand detection using TFBGs. Particularly, spectral position of the TFBG stop band can be precisely determined from the $\pi$ jumps in the phase of the transmitted light. Additionally, outside of a stop band, changes in the modal effective refractive index of the TFBG guided mode can be deduced from variations in the phase of transmitted light. For example, light propagating along the grating of length $L$ in the mode with effective refractive index $n_{eff}$ will incur the phase gain of

$$\phi = 2\pi n_{eff} L / \lambda . (2)$$

Assuming that the smallest measurable change in the phase is $\delta\phi_{min} \approx 0.02 \cdot \pi$ [12], then, the smallest measurable change in the effective refractive index of the guided mode will be $\delta n_{eff} = (\delta\phi_{min}/2\pi) \cdot \lambda/L \Box 10^{-3} RIU$ (we assumed a sensor of $L=17$ mm operated at ~0.3 THz).

We now compare sensitivities of the phase interrogation technique with that of a spectral interrogation technique. Particularly, detection of changes in the spectral position of the TFBG stop band allows direct measurement of the relative changes in the modal refractive index. Indeed, from (1) it follows that the relative change in the center frequency of a TFBG stop band is directly related to the relative change in the modal refractive index as

$$\delta n_{eff} / n_{eff} = -\delta v_B / v_B . (3)$$

From (3) we can now estimate resolution of our sensor. Thus, for a TFBG operating at $v_B \Box 0.375$ THz with $n_{eff} \sim 1$, and a spectral resolution of a THz TDS setup of $\delta v_B \Box 1.5$ GHz, resolution of a TFBG sensor will be $\delta n_{eff} \approx 4 \cdot 10^{-3} RIU$, which is somewhat inferior to the resolution achieved using phase interrogation. At the same time, spectral interrogation offers simplicity of implementation and lack of ambiguity in the result interpretation, which is frequently the case with the phase sensitive methods.

Ultimately, we believe that TFBG represents a convenient technological platform for combining all three interrogation modalities (spectral, amplitude and phase) in the same sensor in order to realize a reliable multi-measurand sensor system with redundant data streams.

In conclusion, TFBGs with 48 periods have been inscribed on subwavelength step-index fibers using $CO_2$ laser engraving. Resulting THz fiber Bragg gratings feature ~4 GHz-wide stop band and ~15 dB transmission loss in the middle of a stop band. Thus created TFBGs were then applied to detection of changes in the thickness of a paper stack that was placed in direct contact with a grating. Experimental spectral sensitivities of ~-0.67 GHz / 10 μm were demonstrated. With a current THz setup that offers 1.5 GHz resolution (600ps traces), we can, thus, detect ~20 μm variations in the paper thickness.

Funding for this work came from the NSERC strategic grant 430420.